# ENVIRONMENT BASED SECURE TRANSFER OF DATA IN WIRELESS SENSOR NETWORKS

Mrs. B. Vidhya[1], Mrs. Mary Joseph[2], Mr. D. Rajini Girinath[3], Ms. A. Malathi[4]

[1]PG Student, Dept of CSE, Anand Institute of Higher Technology, Chennai, Tamil Nadu, India

[2]Assistant Professor, Dept of CSE, Anand Institute of Higher Technology, Chennai, Tamil Nadu, India.

[3]HOD & Prof, Dept of CSE, Anand Institute of Higher Technology, Chennai, Tamil Nadu, India

[4]Assistant Professor, Dept of CSE, Anand Institute of Higher Technology, Chennai, Tamil Nadu, India

*ABSTRACT*

*Most critical sensor readings (Top-k Monitoring) in environment monitoring system are important to many wireless sensor applications. In such applications, sensor nodes transmit the data continuously for a specific time period to the storage nodes. It is responsible for transferring the received results to the Authority on Top-k Query request from them. Dummy data's were added into the original text data to secure the data against adversary in case of hacking the sensor and storage nodes. If storage node gets hacked by adversary, false details will be sent to the authority. An effective technique named aggregate signature to validate the source of the message and also to protect the data against latest security attacks, cryptography technique combined with steganography has been introduced. Indexed based scheme for the database access has also been proposed, to validate the resources against availability before forwarding the data fetch request to storage nodes from Authority.*

*KEYWORDS*

*Top-k Monitoring, Aggregate signature, Cryptography with Steganography, Indexed scheme, Periodic Verification*

## 1. INTRODUCTION

Wireless Sensor Networks focussed the work towards sensing the data by deploying it in the various location to monitor the status of the environment, to report in case of any accident occurs, to investigate the details in military sectors. Depending on each sector, WSN sense and produce the data correctly. The results which is considered to be most critical is said to be as "Top-k Query"[1]. Top-k query fetches the highest sensor data readings among all the sensor node details which are mostly used in many areas. Example:

In Habitat Monitoring, Wildfire spreads across different locations fast in case of temperature and wind gets increase. Clients issue top-k query request to find out which top areas are mostly affected and to take immediate action on this particular area first to prevent the damages to occur more. In Environmental Monitoring, Clients issue top-k query request to identify the areas where the pollution indexes is very high to take considerable action on those critical areas first. Generally attacks can be classified as active and passive attacks. Adversary changes the data which is getting communicated into the communication medium is known as "Active Attack". Adversary monitors the communication medium to track the secure details transferring between the source and destination is known as Passive attack. WSN is prone to many security attacks apart from the one mentioned above. Sensor nodes if gets hacked by the adversary, entire information regarding the critical data's will get exposed. Security for transferring the top-k data in WSN is essential for the data to reach the destination in a secure medium [5].

Efficient security mechanism needs to be implemented to fetch/store the data's in WSN. To obtain the critical data's from the wireless sensor networks top-k query processing has been introduced. The most commonly used techniques in preference-based queries are the top-k query. In order to fetch the top most records from the set of whole records, ranking function needs to be employed. Upon ranking the records based on certain attributes, sorting function needs to be employed to sort it in descending order. Once sorted, the details will be grouped together and first top-k results will be returned to the user. The main advantage of this mechanism is, clients can able to take a stringent action to prevent the damage in becoming worse [2]. Top-k query mechanism is widely employed in distributed systems where the data will get accumulated from different databases and users are often overwhelmed by the variety of similar data. Top-k query fetches the records which clients are interested in and hence the data's can be easily interpreted according to the user purpose. Clients/users employ different ranking functions according to their functionality to retrieve the desired results from it. The applications which get benefit from it includes environment monitoring, habitat monitoring and digital library. Due to more data's get distributed in various locations, top-k query [7] is efficiently used in processing of records in distributed systems.

## 1.1 Security Requirements

### 1.1.1 Mutual Authentication

In Mutual Authentication, two parties needs to register and aware of themselves. In client server technology, both the client and server authenticate themselves initially for assurance activity to establish a communication later. Web to client authentication is also referred to as mutual authentication. Business to Business organization requires extra level of security since the information transferred between them is highly confidential and mostly financial transactions exchanged between the organizations needs to be safeguarded with high security.

### 1.1.2 Confidentiality

Confidentiality is similar to the requirement of privacy. Sensitive information should not be exchanged to the parties other than those who were involved in the communication medium. Hence Access needs to be permitted for accessing the sensitive information should be permitted only for those who were highly authenticated. By encrypting the data, the sensitive information will not be revealed to un-authorized users and it is a common way of ensuring the data confidentiality.

### 1.1.3 Nonrepudiation of Service

User should not deny of sending/receiving the messages which they sent to the corresponding user. In mobile communication system, user should not deny the charges incurred by using the service. Non repudiation can be assured by using the concept of digital signatures, timestamps and confirmation of services.

### 1.1.4 User Anonymity

In wireless communication systems, some clients would not like to disclose their identity and location to third parties. An authentication protocol has been created in wireless communication medium, where each user has been provided with the temporary identity when they get certificate from the Certificate Authority. Then, the assigned identities can be used in that domain.

### 1.2 Existing Works on Data Security

To secure the data against adversary, several secure mechanisms need to be employed while transferring the data in the communication medium. Two approaches carried forward in securing the data in the existing technology were additional evidence and cross check. Additional evidence [1] generates message digest to verify the data. The sender will message digest the data and send to the receiver. Receiver is responsible for verifying the message digest by calculating the new digest for the corresponding received message and to compare the received digest with the newly calculated one. If there are any discrepancies, it is meant that the adversary has modified the data in middle of the communication medium. Crosscheck mechanism [2] distributes the data to the neighbouring sensor nodes. Owner can verify the query result completeness by comparing the data it has received from the sensor node with the same data it has requested from the neighbouring sensor node to validate the received data. Hybrid approach [6] is a combination of above mentioned approaches. It is used to identify the validity of the query result completeness. Verifiable query processing approach employs a mechanism by transferring cryptographic one-way hashes to the intermediate node namely storage node even when they do not have satisfy readings.

MoteSec-Aware security mechanism [5] has been implemented by focussing the works towards secure network protocol and data access control. To detect the replay and jamming attacks in the communication medium, incremental counter has been employed within the internal functionality to count the messages which are getting transferred between the source and destination using symmetric key cryptography in OCB mode. By considering authorization perspective, Key-Lock [9] matching method has been proposed to effectively handle the communication in a secure medium.Continuous top-k query [6] has been proposed as a mechanism to return the most updated results to the users at any point of time. To support the process of the above, close dominance graph technique [4] has been employed to process the data of continuous top-k query. CDG [3] has been used as an indexing scheme for the above approach to effectively retrieve the required data's from the whole set of records.To obtain the temporal variation of aggregate information in sensor applications, continuous aggregation mechanism [7] has been employed. Adversary still can able to modify the information to false temporal variation patterns by changing the series of aggregation results by compromising the sensor and storage nodes.

Sampling based approach [1] has been proposed to verify the validity of the results by picking the random sample of nodes for verification and the verification technique is not relying on any particular in-network aggregation protocol. Verifiable random sampling approach [9] has been proposed to secure the legitimacy of the sampled nodes considered for the previous functionality. To secure the validity of the sample readings, local authentication based on spatial correlation approach [2] has been proposed and effectively used in the above functionality. Information brokering systems [5] has been implemented as a mechanism to interconnect federated data sources via a brokering overlay. Brokers are responsible for taking the routing decisions and to direct client queries to the corresponding data servers. Adopting server-side access control for data confidentiality is not sufficient in the existing IBS [1] since adversary can still infer the details of the data which is getting transferred between the brokers. Hence the secure transfer mechanism needs to be implemented for the data communication across the brokers as well. Attribute correlation attacks [5] and inference attacks has been handled by the mechanism automaton segmentation and query segment encryption to protect the data while transferring in the communication medium as well as to take the routing decision responsibility among a selected set of brokering servers.

Large sensor networks were following the two-tier architecture where the sensor nodes transmit the data to the master node which is responsible for transferring the sensor data to the authority. Securing the data in the two-tier architecture is very essential because of the criticality of the data's. Verifiable Top-k Query mechanism [6] has been implemented to detect any incorrect results returned to the owner by the master nodes in case of the details get hacked by the adversary. Verification is done at the owner end by incorporating and embedding some relationships between the data items. Query conversion scheme has been proposed as a solution to detect the compromise of sensor nodes resulting in transmit of the false details to the master nodes. Random probing scheme has been implemented to detect possible colluding attack from compromised sensor and master nodes. Owner verifies the data which it received by randomly verifying the result among the neighbouring sensor nodes to check against the validity of the result. A light weight scheme RW has been implemented to detect the compromise of sensor node by examining the testimonies from witness nodes. By implementing the above approaches, owner can able to verify the integrity and completeness of the query result verification in the tiered sensor networks

### 1.3 Efficiency and Security Gap

Despite the prior works, following concerns needs to be addressed which were as follows:
- Hybrid Method [7] results in $O(n^2)$ *communications.*
- Motesec-Aware scheme [5] is not applicable to large scale networks. Use of symmetric cryptography in KLM scheme is less efficient, as same key is used across the leader and sensor nodes for data encryption.
- Close Dominance Graph [3] for processing continuous top-k query are designed to compute in a homogenous environment. On query request from multiple clients regarding the same data, frequently raised queries too need to be searched in DB repeatedly for data fetching process.
- Adversary can easily detect the symmetric key [6] shared by all the sensor nodes once any one of the sensor node gets compromised. The verification process at each sensor node has been increased greatly due to the additional verification of neighbouring sampled node.

### 1.4 Contributions

Security mechanisms have been proposed for effective communication in the wireless sensor networks. In particular the following mechanisms have been contributed:

- Elliptic Curve Cryptography has been used effectively to achieve the encryption of the data with the minimal key size.

International Journal of Security, Privacy and Trust Management (IJSPTM) Vol 4, No 1, February 2015

- A Concept named Aggregate Signature has been implemented to achieve the verification of the user authenticity.
- Cryptography technique combined with steganography has been introduced to transmit the data via secure communication channel from storage node to authority.

## 2. PROPOSED MODEL

The diagram depicted below describes the overall architecture diagram for achieving the secure transmission of most critical data in environment monitoring system. Secure mechanisms were adapted in the proposed system in addition to encryption, message digest and dummy reading based techniques which are used in the existing system for transferring the environment data.

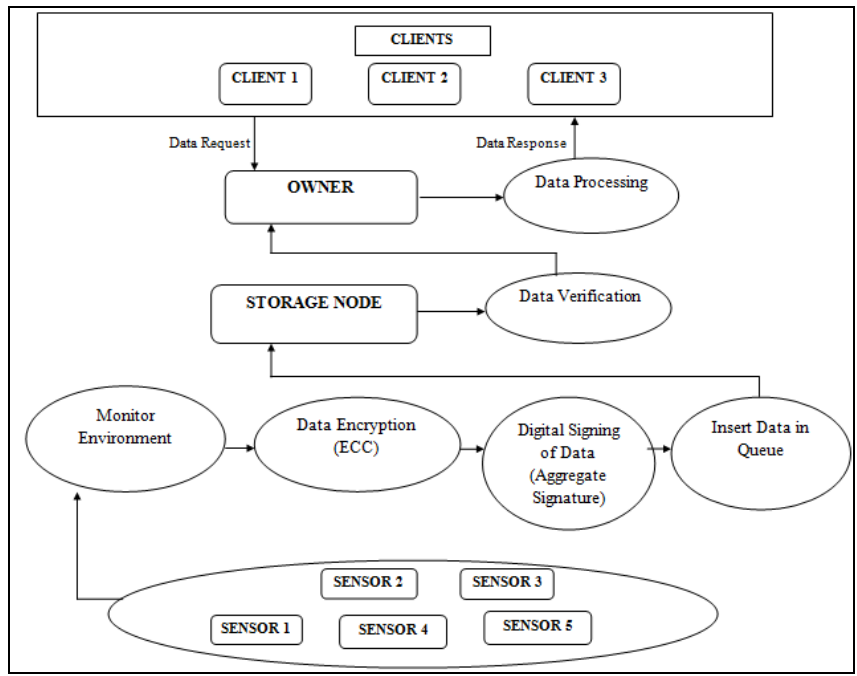

Figure 1: Top-k Data Verification Overall Process Diagram

Proposed system handles the secure communication of transferring the top-k data to clients using asymmetric encryption. To handle the false incrimination of sensor nodes by adversary, aggregate signature has been proposed. The sensor node authenticity has been proven by the above technique. Sensor nodes transfer the signature to the storage node for verification. Storage node is responsible for transferring the data within an image to authority. Thus making the data transfer more secure as shown in Fig 1. Authority upon receiving the data from the storage node, decrypt and stores it in the database before transferring the data to clients. Clients also send their request to authority to fetch the top-k environment data. Upon receiving the request, authority pre-process the data and if needed fetches the result from database. Authority populated the sensor data into the database by processing the steganograph data which has encrypted content within it, which has been sent by the storage node. Sensor nodes are responsible for forwarding the collected data, and the forwarded data is encrypted and the aggregate signature has been generated and sent to the storage node. Verification of validity of aggregate signature will be performed and reported to authority by the storage node in case of invalid signature.

## 3. PROPOSED TECHNIQUES

### 3.1 Sensor Node Data Process

Authority is responsible for deploying the sensor nodes in different locations to monitor the environment and also to respond to the query request from clients. On collecting the required data from environment, sensor node transfers it to the storage node. The environment details will be captured in the image format (JPEG). By using the aggregate signature, the signature will get generate for each of the environment detail captured and it is sent to the storage node for further processing of the signature details. Aggregate signature is a mechanism where the entire sensor signature is grouped into a location and it is verified against each sensor signature.

Signature needs to be appended with the original file to ensure that the data sent by the sensor node is from the valid sensor and not from the compromised sensor node. Compromised sensor nodes pose incorrect hashes as well as results to the authority. In case of such issues, the details of the entire sensor nodes will be hacked by the adversary. Hence to protect the data against different attacks, each sensor node deployed into the different location needs to append signature into it. The signature of all the sensor nodes will be grouped to form a signature called aggregate signature which is then transmitted to the storage node for verification.

### 3.2 Storage Node Verification Process

Processing the signature complexity has been simplified at the storage node by grouping all signatures into one. On collecting the information from the sensor nodes, storage nodes verify the digital signature embedded by the sensor nodes to verify the authenticity of the data sent by the nodes. If the transmitted data is not the reliable sensor node, immediate message transfer of adversary details is sent to the authority for the stringent action to be taken. On successful processing of all the signatures storage nodes decide to transmit the text data to owner by taking a decision on viewing the image received from the sensor node. Dummy data's were introduced into the original content to confuse the adversary from hacking the original content. The content will get translated into the ASCII numbers and then the dummy data's will get inserted into it and the whole data is encrypted (ECC) using asymmetric encryption. To maintain the order of the sentence used in encryption, delimiter has been maintained to uniquely identify the separation of words in between the sentence. The delimiter will also get converted into the ASCII numbers.

### 3.3 Storage Node Data Conversion Process

#### 3.3.1 Message Digest Process

The proposed work considers hashing algorithm for message digest which is HMAC. It is used to message digest the data. The data used will be digested according to the requirements implemented in the algorithm and the sender used to digest the data using the specific unique code and thus the receiver can verify the originality of the message by verifying the content it has received by computing the same message digest code. If both the message digest gets equal, the receiver can ensure that the data sent by the sender has not been modified by the adversary. The concept of the encryption along with message digest scheme achieves query result integrity and completeness verification in the tiered sensor networks.

### 3.3.2 Steganograph Conversion Process

Transformation via steganography results in hiding the original contents sent by the sensor nodes within an image. The idea behind the data transformation is to make the adversary unaware of the transformation of critical data. In the proposed work, cryptography combined with steganography has been achieved in establishing the secure communication between the sensor, storage and MN. The data communication between the storage and master node is also essential and to secure the medium among those is required in transferring the critical data's successfully. Storage node transmits the steganograph data to the authority on successful processing of the sensor node signature. Transferring the data via steganography achieves a great advantage in transferring the secure data in the communication medium. Plainly encrypted messages are visible to all even though the content embedded within it is not visible. Using many technologies, there are possibilities to break the encrypted content by the adversary in trying to get the data transfer in between the communication medium.

Hence steganography came into picture by embedding the details within an image to minimize the possibilities of identifying the content travelling into the message between the concerned parties. Embedding the encrypted content within an image achieves in establishing the data transfer in a more secure manner and adversary will never be knowing that, the details which is getting transferred is critical since it is embedded within an image. Adversary is aware of the image which is getting transferred and not the original content which is embedded within an image.

### 3.4 Authority Data Validation Process

Authority is responsible to retrieve the data embedded within an image using the appropriate steganography decrypting methodologies. Once after fetching the data's the content needs to be decrypted using asymmetric key encryption. Authority and the sensor nodes use different pair of keys to encrypt and to decrypt the data. By using the asymmetric key encryption, the message is encrypted by sender public key. No one will be able to decrypt the content since the owner is the one who will be having the private key to decrypt the content. Adversary will never be able to hack the data since getting the key for decrypting the content is difficult. By using the above approach, confidentiality is maintained in which the message signed by the sender can only be viewed by the person who has the corresponding decrypting key. The decryption key will be shared by the owner only to the trusted parties.

With the help of message digesting the data in asymmetric key, second level of assurance of data is also has been ensured that the data content sent by the sender is unmodified and thus results in achieving the result integrity and completeness while transferring the data in the communication medium. Upon decrypting the data, authority is responsible for updating the corresponding sensor details into the database. All the details regarding the sensor environment information will be stored into the database corresponding to each sensor id by maintaining it has a private key in DB.

### 3.5 Client Data Fetching Process

On receiving the client request on top-k query, authority is responsible for fetching the corresponding request and to transfer the data to the clients. To perform such functionalities, two approaches needs to be followed namely sorting and ranking. Ranking functionality needs to be considered by considering the data's and to rank correctly according to their percentage of monitored data regarding environment. According to their attributes, the data will get ranked. Sorting functionality which is based on the ranking attribute, sort the details into the descending order in order to fetch the details of the environment as expected by the user. Once the above two functionalities has been performed, the requested client top-k data will be returned to the

International Journal of Security, Privacy and Trust Management (IJSPTM) Vol 4, No 1, February 2015

users/clients who are responsible for taking the stringent action on verifying the severity level of the data.

Table 1. Notation Table

| Symbol | Abbreviation |
|---|---|
| HMAC | Hash Message Authentication Code |
| WSN | Wireless Sensor Network |
| MN | Master Node |
| STN | Storage Node |
| SN | Sensor Node |
| ECC | Elliptic Curve Cryptography |
| $\Omega_k$ | Result of Top-k query |
| k | No of records |

## 4. ELLIPTIC CURVE ENCRYPTION MECHANISM

Securing critical data transfer about the environment in military applications are vital since such data assist in decision making process. The secured approaches employed for transferring the top-k query result in the proposed work assist in transferring the critical data's more secure in the multitier architecture of WSN. Elliptic Curve encryption mechanism has been implemented to encrypt the data using minimal key size.

### 4.1 Basic Idea of ECC

Elliptic Curve Cryptography depends on three variables
- Prime Number to be selected as maximum
- A Curve equation
- Public point on the curve

**Private Key:** Chosen *Private number*
**Public Key:** *Public point* dotted with itself *Private number* times
Calculating the private key from the public key in this kind of system is said to be elliptic curve discrete logarithm function. Elliptic curve is based on the points which should satisfy the equation.
$y^2 = x^3 + ax + b$
The graph for the above equation with the points plotted will look similar to the below:

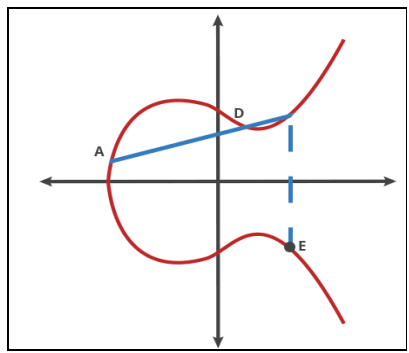

Figure 2. Elliptic Curve Cryptography graph

Points can be plotted in the graph to get a new point. Any two points can be dotted together to get a new point as follows:

A dot B = C
A dot C = D
A dot D = E

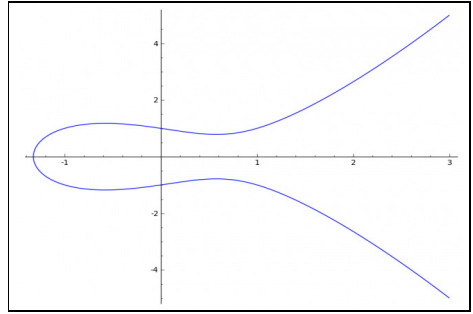

Figure 3. Example curve for the equation $y^2 = x^3 - x + 1$

## 4.2 Algorithmic Description of ECC

### 4.2.1 Key Generation

Public and private key will get generate in key generation part. Receiver Public key is used by the sender to encrypt the sender message and the receiver will use the private key to decrypt the message. Select a number **'c'** within the range of **'n'**. Following equation is used to generate the public key.

*Q=c\*X*
**c** = Random number that we have selected within the range of (**1 to n-1**).
**X** =Point on the curve.
**Q**=Public Key.
**c**= Private Key.

### 4.2.2 Encryption

Sender is transferring the message 'm' to the receiver. The message needs to be represented in the elliptic curve. Consider the message 'm' has the point 'M' on the curve 'E'. The number 'k' needs to be randomly selected from [1 – (n-1)]. Encryption process will generate two cipher texts $Ci_1$ and $Ci_2$.

*$Ci_1$=k \* X*
*$Ci_2$= M + k\*Q*
The points *$Ci_1$* and *$Ci_2$* will be sent to the receiver.

### 4.2.3 Decryption

The original message will be retrieved by decrypting the content which can be done using the below equation.

*M=$Ci_2$ – c \* $Ci_1$*
The message *'M'* is the original message which the sender transferred to the receiver.

**4.2.4 Proof**

*M=$Ci_2$ – c * $Ci_1$*
'M' can be represented as '$Ci2$ – c * $Ci1$'
$Ci_2$ – c * $Ci_1$= (M + k * Q) – d * (k * X)     ($Ci_2$ = M + k * Q and $Ci_1$ = k * X)
= M + k * d * X – d * k *X                (cancelling out k * d * X)
= M (Original Message).

## 5. PERFORMANCE ANALYSIS

There are many asymmetric algorithms used for encryption. Symbols are either permuted or substituted in symmetric key cryptography but numbers are manipulated in asymmetric cryptography. In the proposed system, the concepts of Elliptic Curve Cryptography have been used. In Public Key cryptography, Elliptic Curve Algorithm is the standardized IEEE format. ECC offers equal security compared to RSA with small key size.

The following graphs summarize the outcomes of the experiments carried out during this study. Table 2 shows the different algorithm sizes compared with the ECC implementation. ECC offers equal security compared to RSA with small key size.

Table 2. Algorithm Key Sizes

| S.No | Symmetric Key Size (in Bits) | RSA Key Size (In Bits) | ECC Key Size (in Bits) |
|---|---|---|---|
| 1 | 80 | 1024 | 160-223 |
| 2 | 112 | 2048 | 224-255 |
| 3 | 128 | 3072 | 256-383 |
| 4 | 192 | 7680 | 384-511 |
| 5 | 256 | 15360 | 512+ |

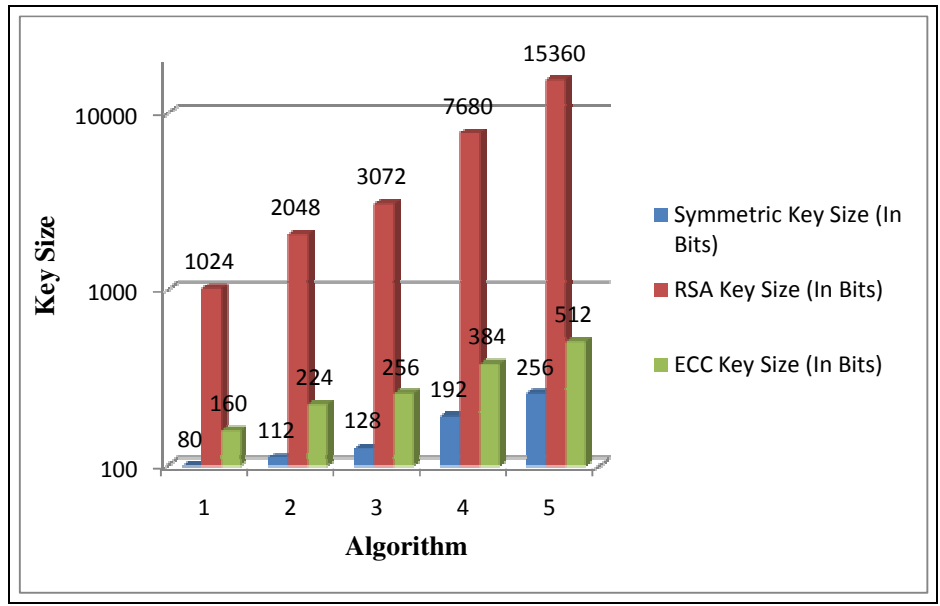

Figure 4.  Comparison Graph for Algorithm Key Sizes

## 6. CONCLUSION

Securing critical data transfer about the environment in military applications are vital since such data assist in decision making process. Recent secure approaches needs to be adhered for transferring the data. The secured approaches employed for transferring the top-k query result in the proposed work assist in transferring the critical data's more secure in the multitier architecture of WSN. It thus results in achieving the top-k query result integrity and completeness verification which is essential in many sensor related applications namely industrial civilian areas, habitat monitoring and environment monitoring.

## REFERENCES

[1] Chia-Mu Yu, Guo-Kai Ni, Ing-Yi Chen, Erol Gelenbe & Sy-Yen Kuo, (2014) "Top-K Query Result Completeness Verification In Tiered Sensor Networks," Ieee Transactions On Information Forensics Security, Vol. 9, No. 1, Pp. 109-123.
[2] Yao-Tung Tsou, Chun-Shien Lu & Sy-Yen Kuo, (2013)"Motesec-Aware: A Practical Secure Mechanism For Wireless Sensor Networks", Ieee Transactions On Wireless Communications, Vol 12, No 6, Pp.2818-2822.
[3] Bagus Jati Santoso & Ge-Ming Chiu, (2014) "Close Dominance Graph: An Efficient Framework For Answering Continuous Top-K Dominating Queries", Ieee Transactions On Knowledge And Data Engineering, Vol 26, No 8, Pp.1854-1864.
[4] Lei Yu, Jianzhong Li, Siyao Cheng, Shuguang Xiong & Haiying Shen, (2014) "Secure Continuous Aggregation In Wireless Networks", Ieee Transactions On Parallel And Distributed Systems, Vol 25, No 3, Pp.763-773.
[5] Fengjun Li, Bo Luo, Peng Liu, Dongwon Lee & Chao Hsien Chu, (2013) "Enforcing Secure And Privacy-Preserving Information Brokering In Distributed Information Sharing",Ieee Transactions On Information Forensics And Security, Vol 8, No 6, Pp. 889-895.
[6] Rui Zhang, Jing Shi, Yanchao Zhang & Xiaoxia Huang, (2014) "Secure Top-K Query Processing In Unattended Tiered Sensor Networks",Ieee Communication And Information System, Huazhong University Of Science And Technology, Vol 25, No 3, Pp. 763-773.
[7] Daojing He, Sammy Chan & Shaohua Tang, (2014) "A Novel And Lightweight System To Secure Wireless Medical Sensor Networks",Ieee Journal Of Biomedical And Health Informatics, Vol. 18, No. 1, Pp. 317-324.
[8] Mohamed M.E.A. Mahmoud, Sanaa Taha, Jelena Misic & Xuemin (Sherman) Shen, (2014) "Lightweight Privacy-Preserving And Secure Communication Protocol For Hybrid Ad Hoc Wireless Networks", Ieee Transactions On Parallel And Distributed Systems, Vol. 25, No. 8, Pp. 2078-2088.
[9] Emiliano De Cristofaro & Roberto Di Pietro, (2013) "Adversaries And Countermeasures In Privacy-Enhanced Urban Sensing Systems", Ieee Systems Journal, Vol. 7, No. 2, Pp. 312-320.
[10] Omar Hasan, Lionel Brunie, Elisa Bertino & Ning Shang, (2013) "A Decentralized Privacy Preserving Reputation Protocol For The Malicious Adversarial Model", Ieee Transactions On Information Forensics And Security, Vol. 8, No. 6, Pp. 950-960.

International Journal of Security, Privacy and Trust Management (IJSPTM) Vol 4, No 1, February 2015

## AUTHORS

**Mrs. B. Vidhya** received the bachelor's degree in 2009 in Information technology from Anna University in Vel Multi Tech SRS Engineering College. She is currently doing Master's degree in computer science & engineering from Anna University in Anand Institute of Higher Technology. Her research areas are wireless sensor networks, database technology, and network security.

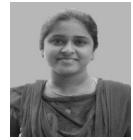

**Mrs. Mary Joseph** is working as an Assistant Professor, Department of Computer Science and Engineering, Anand Institute of Higher Technology, Chennai, Tamilnadu, India. She has completed B.E Computer Science and Engineering in Thangavelu Engineering College, Chennai in the year 2007 and completed M.E Computer Science and Engineering in M.N.M Jain Engineering College, Chennai in the year 2010. Her research areas are data mining and computer networks.

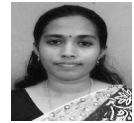

**Mr. D. Rajinigirinath** received the B.E. degree in Computer Science and Engg from Madras University, M.Tech. degree in Computer Science and Ph.D from Anna University, Chennai in the area of Information and Communication Engineering. His research interest includes Computer Networks, Adhoc Networking, Mobile Computing, Artificial Intelligence and Image Processing. He has presented 10 papers in International conference and 25 papers in National Conference. He has Published 25 Papers in International and National Journals. Presently working as Professor and Head of the Department of Computer science and Engineering in Anand
Institute of Higher Technology, Chennai.

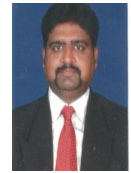

**Ms. A. Malathi** is working as an Assistant Professor, Department of Computer Science and Engineering, Anand Institute of Higher Technology, Chennai, Tamilnadu, India. She has completed B.Tech Information Technology in Easwari Engineering College, Chennai in the year 2004 and completed M.E Computer Science and Engineering in SRM University, Chennai in the year 2013. Her research areas are computer networks and network security.

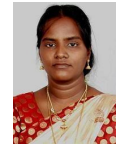